\documentclass[aps,nofootinbib,twocolumn,10pt]{revtex4}
\usepackage{times}
\usepackage{graphicx}
\usepackage{amsmath}
\usepackage{epsfig}
\usepackage{amssymb,amsfonts,nicefrac}

\def\zi{\mathbb{Z}}

\def\slr{$SL(2,\mathbb{R})\ $}
\def\slc{$SL(2,\mathbb{R})/U(1)\ $}
\def\sla{$\widehat{\mathfrak{sl}} (2,\mathbb{R})$}

\begin{document}
\onecolumngrid
\begin{flushright}
     hep-th/0502101
\end{flushright}
\title{Strings and D-branes in holographic backgrounds}
\preprint{xxx}
\author{Dan Isra\"el\footnote{e-mail: {\tt israeld@phys.huji.ac.il}}} 
\affiliation{The Racah Institute of Physics, 
The Hebrew University of Jerusalem, 91904 Jerusalem, Israel   }

\begin{abstract}
We review recent progress in the study of 
non-rational (boundary) conformal field theories and their applications 
to describe exact holographic backgrounds in superstring theory. 
We focus mainly on the example of the supersymmetric 
coset \slc, corresponding to the two-dimensional black hole, 
and its dual N=2 Liouville. 
In particular we discuss the modular properties of their 
characters, their partition function as well as the exact 
boundary states for their various D-branes. Then these 
results are used to construct the corresponding quantities 
in the CFT of the NS5-brane background, with applications 
to Little String Theories.  
\vskip.2cm
{\it From talks given in the RTN2004 meeting, Kolymbari and 
Joint HEP Israeli Seminar, Oct. 2004}
\end{abstract}
\maketitle                  

\section{Introduction}
Conformal field theories are the natural 
building blocks for the exact perturbative description of 
superstring theory around non-trivial backgrounds. It 
is possible to describe a large class of solutions,   
albeit without Ramond-Ramond fluxes. 
In addition to  non-trivial compactifications and supersymmetric curved 
vacua, some exact time-dependent solutions are known. 
However for the two latter categories, the conformal field theories 
that we shall need are {\it non-rational conformal field theories} 
(\textsc{nrcft}s). For technical 
and conceptual reasons they are far more difficult to handle 
that their rational cousins (\textsc{rcft}s), and the simplest examples 
have been fairly understood only in the recent years. 

This progress has been largely motivated by their relevance 
to describe backgrounds with an explicit 
implementation of holography. The only examples with only 
Neveu-Schwarz--Neveu-Schwarz fluxes are generated by fundamental strings 
and NS5-branes sources, the most famous one being the 
three-dimensional anti-de Sitter 
spacetime\cite{Balog:1988jb,Maldacena:2000hw}, whose holographic nature 
is clearly established\cite{Maldacena:1997re}. 
However the holographic nature of five-branes 
solutions\cite{Callan:1991dj,Kounnas:1990ud} is less clear, 
mainly because of the non-trivial nature of the dual non-gravitational 
theory, called {\it Little String Theory}\cite{Seiberg:1997zk,Aharony:1998ub}. This type 
of holography with a linear-dilaton background can be generalized e.g. to 
sub-critical superstrings models\cite{Giveon:1999zm}, 
in particular to the two-dimensional case, where the tools of matrix 
theory are very powerful to handle the theory\cite{Douglas:2003up}. 

These constructions can be recast as non-rational analogues of 
Gepner models\cite{Gepner:1987qi}, where the building blocks are, for the compact 
part, Kazama-Suzuki N=2 cosets and  for the non-compact part 
the super-cosets \slc\cite{Eguchi:2000tc,Mizoguchi:2000kk,Yamaguchi:2000dz,Eguchi:2003yy,Israel:2004ir,Eguchi:2004ik}. 
Adding fundamental strings amounts to lift one of those non-compact cosets to \slr, i.e. 
three-dimensional anti-de Sitter space\cite{Giveon:1999jg}. 
Thus the main non-trivial building block for these constructions is the 
(super) coset \slc\cite{Elitzur:1991cb,Mandal:1991tz,Witten:1991yr}
that we shall study in detail in this note. 
Below we shall begin with some general features of \textsc{nrcft} that 
can be inferred from the solved theories of this kind. Section two 
deals with the the closed string sector of the 
super-coset \slc as well as its mirror N=2 Liouville. 
In section three we consider the exact construction of D-branes in this model. Finally 
in section four we lift these results to the background of NS5-branes. 

\section{General aspects of non-rational conformal field theories}
The non-rational \textsc{cft}s are of course defined in opposition 
to rational ones. The latter are characterized by having 
a finite number of primaries of their chiral algebra. Powerful 
algebraic techniques are available to solve them completely, 
with or without boundary\footnote{although only a subset of D-branes 
is known in general.}. Famous examples 
of these models are minimal models of the (super)conformal algebra, 
\textsc{wzw} models on compact groups and their cosets.  

In \textsc{nrcft} the aforementioned techniques are no longer useful 
or need to be adapted. The complications arise first because the 
Hilbert space is now constructed out of an {\it infinite} number of 
primary fields, that may even form a continuum of states, since these models 
have non-compact target spaces. In general the spectrum of these theories
splits into two categories of states. On the one hand the spectrum contains 
{\it continuous representations}, corresponding to asymptotic states 
propagating in the non-compact ``radial'' direction(s). On the 
other hand the states of the {\it discrete representations} correspond 
to (a finite set of) localized bound states.  
As we may infer from general considerations of scattering theory, these 
two kinds of states mix. It is known for example 
in the coset \slc that the reflection amplitude for continuous 
representations has poles whenever it is analytically continued to a discrete 
representation\cite{Giveon:1999px}.
The modular properties of the character of those theories exhibit a 
similar pattern. In the bosonic coset\cite{Israel:2004xj} and its supersymmetric 
version\cite{Eguchi:2003ik,Ahn:2003tt}
the modular transformations of the characters can be represented 
as contour integrals in the momentum plane of the non-compact 
directions, using a lemma proven in\cite{Miki:1989ri,Israel:2004xj}. 
For the transformations of the discrete characters, this integral has to be shifted in 
order to recast the integral as a contribution from continuous
representations. The residues of the poles crossed during this process give 
characters of discrete representations. 
Thus, while the continuous representations modular transform onto 
continuous representations, the discrete representations give both 
discrete and continuous ones. These new qualitative features 
complicate the construction of partition functions and boundary states 
by a large amount. 

These \textsc{nrcft}s can still be solved by the {\it conformal bootstrap}, 
using the chiral symmetries of the model. 
Apart from the pattern described above, 
some features like the absence of identity representation in the 
closed string spectrum complicate the analysis of the 
factorization constraints.  Generalizing the seminal works 
on Liouville theory\cite{Zamolodchikov:1995aa,Teschner:1995yf,Teschner:2001rv}, 
one uses some specific representations that 
are degenerate w.r.t. the chiral algebra. In those cases the operator 
product expansion (\textsc{ope}) with any field is finite --~in contrast with a 
generic \textsc{ope} of these theories~-- and differential equations 
have to be obeyed. An important assumption is that the quantity 
we wish to compute can be {\it analytically continued} to the 
degenerate representations that are not continuous ones. Then, one can 
assumes on general grounds that some quantities involving these 
degenerate fields are perturbatively computable, and adding some 
assumption of strong/weak 
duality on the worldsheet gives the exact solution of the theory. 
However it is sometimes possible to use {\it only} the chiral symmetry 
of the model and the factorization constraints to solve the 
model\cite{Teschner:1997ft,Teschner:1999ug,Teschner:2001gi}.
In this case we are close to tell the same story as for rational theories.    

A last important aspect of \textsc{nrcft} that we wish to highlight is 
that, for specific points in the parameter space, these theories 
simplify and acquire some kind of rational behavior. The simplest 
example is the free \textsc{nrcft} for a boson on a circle. When the 
radius squared turns out to be rational, i.e. $R=\sqrt{2r/s}$, the 
theory has an extended chiral symmetry generated by $J_0 = \imath \partial X$ 
and $J_\pm = \exp (\pm \imath \sqrt{2 rs }X)$. Then by summing over 
the orbits of this symmetry one gets a finite set of {\it extended characters}.  
Intuitively one may guess that this behavior generalizes to the 
(super) coset \slc for rational level, since the target space asymptotes a 
cylinder of radius $\sqrt{2k}$. It is indeed the case\cite{Eguchi:2003ik,Israel:2004xj}
and the properties of the theories simplify. In particular in the supersymmetric case 
one obtain a finite set of rational N=2 R-charges, which is desirable 
to construct space-time supersymmetric vacua.

\section{The supersymmetric cigar and N=2 Liouville}
As proposed in the introduction we shall focus now on the 
super-coset \slc, since it is a prototypic example 
of \textsc{nrcft} and a basic building block of most of the 
non-trivial superstrings vacua. This theory is obtained 
by a straightforward application of the rules of coset 
construction\footnote{
We consider the gauging of the {\it elliptic subgroup} giving the 
Euclidean black hole.} to the supersymmetric \textsc{wzw} model \slr at level $k$. 
For the {\it axial} coset the sigma model is well defined because the action of the 
gauge field has no fixed point, and corresponds to an Euclidean 
two-dimensional black hole. 
The spectrum of primaries is obtained 
by descent from AdS$_3$. It contains both discrete (real spin $j$)
and continuous representations (imaginary spin $j = \nicefrac{1}{2} 
+ \imath s$, $s\in \mathbb{R}$) 
of the affine $\widehat{\mathfrak{sl}} (2,\mathbb{R})$ algebra, the spins of the 
former being restricted to the improved unitary range 
$\nicefrac{1}{2} < j < \nicefrac{k+1}{2}$\cite{Maldacena:2000hw}.
This is confirmed by a computation of the one-loop 
vacuum amplitude\cite{Eguchi:2004yi,Israel:2004ir}, as for the 
bosonic coset\cite{Hanany:2002ev} and the \slr \textsc{wzw} model\cite{Israel:2003ry}.
 
The worldsheet-supersymmetric 
partition function can be computed using the powerful techniques 
of marginal deformations of \textsc{wzw} models\cite{Israel:2003ry}. 
Indeed one can start 
with the supersymmetric \slr model, for which the fermions are free,  
and deform with the truly marginal operator 
$(j^3 + \psi^+ \psi^- ) (\bar{\jmath}^3 - \bar{\psi}^+ \bar{\psi}^- )$ made with the 
total (left and right) elliptic currents of \sla. 
After analytic continuation to 
an Euclidean target space and an infinite deformation along the 
elliptic subgroup, one obtains the desired partition function of the 
supersymmetric coset\cite{Israel:2004ir} (see also\cite{Eguchi:2004yi}). 
This amplitude should split naturally 
into (non-minimal) characters of the N=2 superconformal algebra, 
since it is the largest chiral algebra of the model, and those characters 
appear naturally into the branching relations of the supersymmetric 
coset\cite{Eguchi:2004yi,Israel:2004jt}. 
An exact decomposition of the partition function has been carried out 
in\cite{Israel:2004ir}. We have found first a contribution of discrete representations 
filling the improved unitary range, with the correct multiplicities 
for all the descendants. For the continuous representations, the story 
is a little bit more complicated since their contribution is divergent, 
due to the infinite volume available for them. An infrared regularization 
of the partition function is possible (as in\cite{Maldacena:2000kv,Hanany:2002ev}), 
leading to a finite non-trivial density of continuous representations, 
compatible with N=2 supersymmetry. However this regulator breaks 
(super)conformal symmetry, and there is a price to pay: as follows from 
our exact analysis, the partition function contains an extra 
non-universal contribution which is not related to the N=2 algebra.

The super-coset \slc has been conjectured to be 
dual to another \textsc{cft} with N=2 superconformal symmetry, the N=2 Liouville 
theory\cite{Girardello:1990sh}; this statement is 
the supersymmetric version of the duality between the bosonic 
coset \slc and sine-Liouville theory\cite{FZZ}. In both cases 
it amounts to a strong/weak duality on the worldsheet. Evidence 
for this equivalence comes from a sigma-model mirror
symmetry\cite{Hori:2001ax}, and from the agreement between 
perturbative computations assuming that the duality 
holds\cite{Giveon:2001up} and the conformal bootstrap 
results\cite{Teschner:1997ft,Teschner:1999ug}\footnote{Another 
method is given in\cite{Tong:2003ik}.}. However we 
would like to argue that these results come from a more fundamental 
structure of these theories. Indeed, both theories possess the same 
chiral algebra, which is the (non-minimal) N=2 \textsc{sca}. 
This algebra can be decomposed into the bosonic 
coset \slc and a free 
boson\cite{Dixon:1989cg,Bakas:1991fs,Israel:2004xj}, or from 
a complementary point of view, a non-minimal N=2 algebra 
supplemented by a free time-like boson is lifted to the supersymmetric 
\slr current algebra\cite{Eguchi:2004yi,Israel:2004ir,Israel:2004jt}. 
The conformal bootstrap results for the Euclidean 
AdS$_3$\cite{Teschner:1997ft,Teschner:1999ug} have been obtained 
using {\it only the chiral symmetries} of the model, without 
an explicit reference to a specific action. Thus they can be applied 
to the supersymmetric \slc coset by the coset construction, and 
to the N=2 Liouville theory as well, because they both lift
to the same current algebra\cite{Israel:2004jt}. Therefore the 
theories can differ only by the way the left and right 
representations of the algebra are glued in the closed string 
spectrum. The vector coset \slc has a singular sigma-model, 
therefore it receives substantial 
corrections~; as shown by marginal deformation techniques\cite{Israel:2003ry}, 
its single cover is given by a $\zi_k$ orbifold of the cigar. It is likely 
that the N=2 Liouville theory describes this vector coset, as it 
is suggested by using mirror symmetry\cite{Hori:2001ax}. 
In the following we shall see that all this reasoning about the duality 
extends straightforwardly when one adds a boundary to the \textsc{cft} 
and construct exact D-branes.

\section{Boundary N=2 Liouville from boundary AdS$_3$}
The study of D-branes in these exact superstrings backgrounds 
is essential, in order to understand the non-perturbative 
dynamics in these non-compact manifolds. Moreover 
it may give some indications about the holographic 
degrees of freedom we are looking for. The construction 
of the exact boundary states, which contain all the information 
about the couplings of the D-branes to the closed string states, 
follow the same logic as before. Indeed, after the boundary 
bosonic Liouville theory has been 
solved\cite{Fateev:2000ik,Teschner:2000md,Zamolodchikov:2001ah}, 
the conformal bootstrap methods have been employed successfully 
to construct the D-branes in Euclidean AdS$_3$\cite{Ponsot:2001gt}. 
These results have been used later\cite{Ribault:2003ss} to study the 
bosonic coset \slc. 

In\cite{Israel:2004jt} we constructed the 
D-branes in the super-coset \slc using similar methods. A very 
important aspect of this analysis follows from the fact that 
the conformal bootstrap of\cite{Ponsot:2001gt} uses only 
the chiral symmetries of the models. Therefore the arguments we gave in the previous 
section about the duality super-coset \slc / N=2 Liouville
extend straightforwardly in the presence of a boundary. However 
one should be careful about the way left and right representations 
are glued, in order to construct the basis of 
Ishibashi states\cite{Ishibashi:1988kg}. 
Then the various boundary states of the theory can be constructed 
by descent from D-branes of Euclidean AdS$_3$. Indeed the \textsc{brst} 
formalism allows to rewrite the coset theory as the constrained product 
$SL(2,\mathbb{R})_{k+2} \times \text{Fermions} 
\times U(1)_k \times \text{Ghosts} \times \text{Superghosts}$. 
The boundary state will be a tensor product of boundary states for 
each of these factors, whose boundary conditions are correlated through the 
preserved \textsc{brst} current. 
In particular, there is a direct connection between the boundary conditions 
for the currents of the \sla algebra and the gluing conditions 
of type A or B defined in\cite{Ooguri:1996ck}. 
The D-branes that we obtained satisfy by construction the 
factorization constraints, since they descend from consistent 
D-branes in Euclidean AdS$_3$. However we needed to check 
that the Cardy condition\cite{Cardy:1986ie} --~the consistency 
of the annulus diagram in the open string channel~-- held, since our D-branes 
are constructed out of a non-unitary theory.
\begin{figure}[!ht]
\vskip-1cm
\centering
\includegraphics[width=20mm]{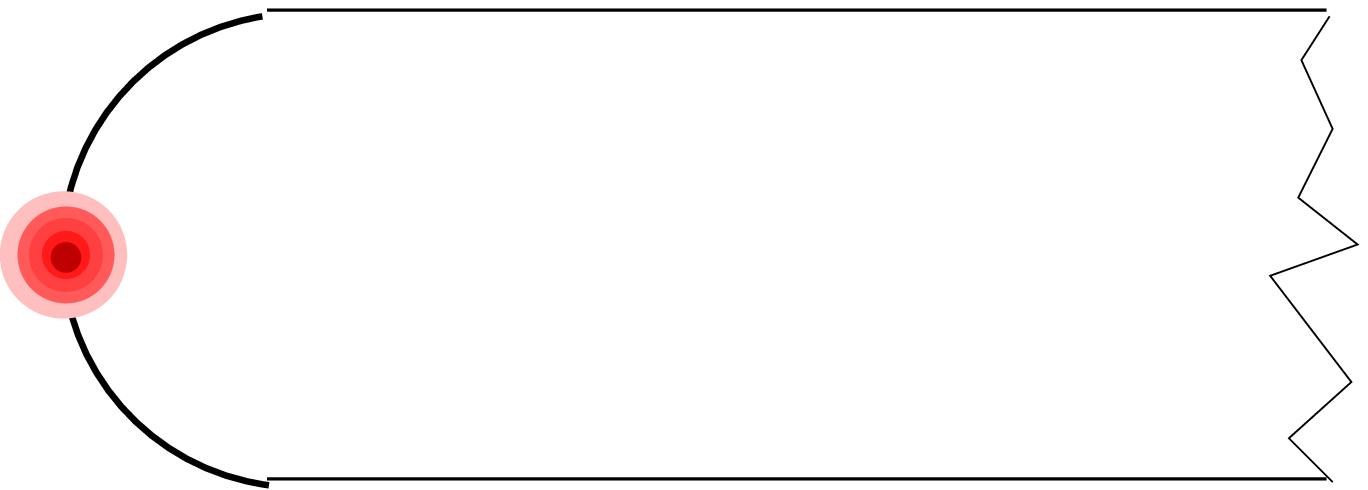}
\hskip5mm
\includegraphics[width=20mm]{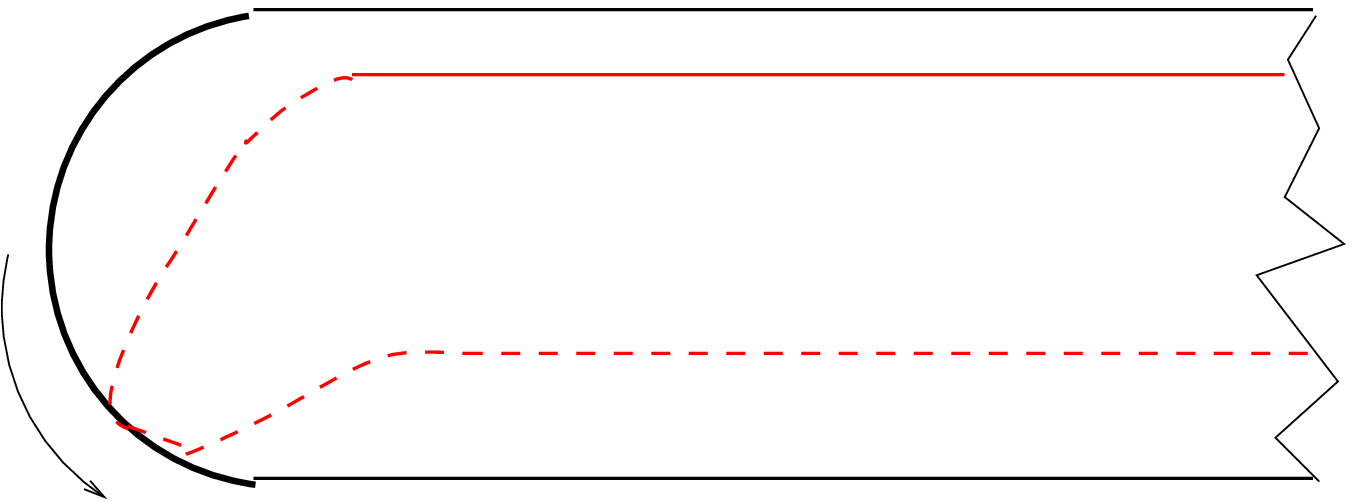}
\includegraphics[width=26mm]{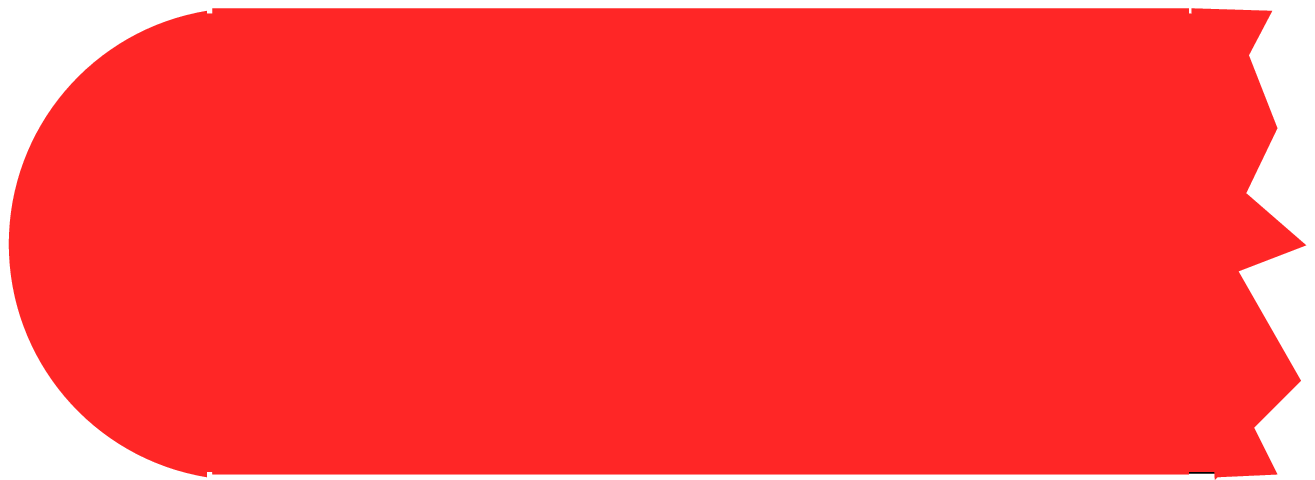}
\caption{The consistent D-branes in the cigar: D0-, D1- and D2-branes 
(left to right)\\}
\label{cigarbranes}
\vskip-.8cm
\end{figure}
The D-branes of the cigar are depicted in fig.~\ref{cigarbranes}. 
\paragraph{A-type branes}
They correspond to the A-type gluing conditions of the N=2 
\textsc{sca}. In the cigar they are D1-branes, extending to infinity 
in the radial direction, with Dirichlet boundary conditions for 
the compact $U(1)$. 
They couple only to closed strings of the continuous representations. The 
annulus amplitude gives a continuous spectrum of open strings, albeit 
with a divergent density since the D-brane is non-compact. We encountered 
previously the same problem with the closed strings partition function. However 
in the present case it can be solved by considering as 
in\cite{Ponsot:2001gt,Ribault:2003ss} the {\it relative partition function} 
w.r.t. a reference brane. The non-universal part of the density is finite 
and related to the boundary reflection amplitude. 

\paragraph{B-type branes}
They correspond to the B-type boundary conditions, and 
can be of two kinds depending of the D-brane of 
Euclidean AdS$_3$ they come from. The first class of B-branes are 
point-like D0-branes, sitting at the tip of the cigar. Their 
boundary states contain couplings both to the discrete and continuous 
representations. They carry an integer label corresponding in the open string channel 
to the spin of a {\it finite representation} of the \sla algebra. 
These representations are non-unitary except the trivial representation, 
thus only the corresponding D0-brane is physical. The 
second kind of B-brane is much more involved. They are D2-branes covering 
all the cigar with a magnetic field on their worldvolume. On general 
grounds we expect that they carry carry a D0-brane charge. There 
is indeed a {\it relative quantization condition} on the 
magnetic field\cite{Ribault:2003ss} allowing to induce a D0-like 
contribution for the open string partition function. However we found 
that they appear with negative multiplicites, hence only the 
D2-brane without magnetic field seems to be physical.  

In related works\cite{Eguchi:2003ik,Fotopoulos:2004ut} the D-branes of the N=2 
Liouville theory and the bosonic \slc were studied using the method called ``modular 
bootstrap'', which assumes that 
the Cardy condition is stringent 
enough to find the correct boundary states, and that 
the Cardy ansatz is valid 
for the \textsc{nrcft}s. While this method gives the same D0-brane, 
it cannot give the D1-branes of the cigar which are not of the same 
type as the D0-brane associated to the trivial representation. However it gives 
other types of extended D-branes. For instance, different D1-branes but they don't seem to be 
consistent with the conformal bootstrap and the semi-classical limit\cite{Fotopoulos:2004ut}. 
A new kind of D2-branes descending from dS$_2$ D-branes of AdS$_3$ 
has been found, and it would be interesting to check their compatibility with the factorization constraints. Some 
other works\cite{Ahn:2003tt,Ahn:2004qb,Hosomichi:2004ph} also apply directly the 
conformal bootstrap method to the N=2 \textsc{sca}. 

\section{Closed and open strings near NS5-branes}
The results given above can be lifted to various superstring setups. 
Here we will focus on the background created by NS5-branes distributed 
on a topologically trivial circle, 
known to be T-dual to an exact \textsc{cft}\cite{Sfetsos:1998xd}. When 
all the NS5-branes are separated from each other, the background 
is expected to be perturbative and one can take a 
{\it double scaling limit}\cite{Giveon:1999px} where gravity decouples and 
this perturbative nature holds. We have shown in\cite{Israel:2004ir} that 
the complicated solution for the ring of five-branes in this limit can 
be obtained as a {\it null gauging} of the super-\textsc{wzw} model $SL(2,\mathbb{R}) 
\times SU(2)$. This \textsc{cft} has an N=4 \textsc{sca} on the worldsheet hence 
there are no perturbative corrections to the effective action. However it is 
likely that instantons corrections show up, and they are indeed captured 
by the supergravity solution. Computing them from the 
worldsheet, following\cite{Tong:2002rq}, is a challenging task. 
Using the \textsc{brst} construction, the null coset can be recast as a $\zi_k$ orbifold 
of \slc$\times SU(2) / U(1)$. It is indeed under this form that the double scaled 
little string theory has been known, using duality\cite{Ooguri:1995wj} and holographic 
arguments\cite{Giveon:1999px}. The $\zi_k$ orbifold can be thought as coming 
from the \textsc{gso} projection generalizing Gepner models. However we would like 
to stress that this orbifold changes deeply the background of the effective theory, 
because in the semi-classical limit its 
twisted sectors become very light; hence the correct geometry is not given 
by the sum of the two coset factors but by the metric of\cite{Sfetsos:1998xd}. 

Once this identification has been understood it has been possible to write the 
one-loop amplitude for this NS5-brane backgroud\cite{Israel:2004ir} 
(see also\cite{Eguchi:2003ik}). The various \textsc{bps} D-branes in this 
background are now under study~\footnote{Part of these D-branes 
have been considered in\cite{Eguchi:2003ik,Eguchi:2004ik,Nakayama:2004yx}
in a related context.}. In particular there are new non-factorizable 
D-branes that can be constructed out of the coset D-branes\cite{Israel:2005fn}. 
The compact D1-branes of type IIB are also especially interesting since as we shown 
they can be related to the matrix model of Little String Theory in the large $k$ 
limit. In type IIA we have also D4-branes stretched between the NS5-branes 
on which a D=4, N=2 \textsc{sym} theory lives\cite{Witten:1997sc}. Quite 
remarkably, the one-point function for these D4-branes can be related to the 
beta-function of the gauge theory. 
 
\section{Conclusions}
It seems now that these new results will allow shortly to solve 
the non-rational conformal field theory as much as their rational cousins. 
Some quantities like the boundary three-point functions are not known in general 
but it is only a matter of computational complexity. Hence a huge class 
of space-time supersymmetric non-compact backgrounds can be studied in 
great detail in the string theory regime, much like the Gepner models some 
years ago. These non-compact models are also related to Lorentzian 
backgrounds such as black holes and cosmological models, and the methods 
presented here can in principle be extended to those cases by some appropriate 
analytic continuation. However it leads to important problems related to the 
choice of the vacuum, and the stability of the Wick-rotated theory is not 
guaranteed. Solving these issues will shed a new light on space-like 
singularities in string theory which are so far not understood. 

\section*{Acknowledgements}
It is a pleasure to thank C.~Kounnas, A.~Pakman and J.~Troost for the collaborations 
leading to the present results. I also thank C.~Bachas, S.~Elitzur, A.~Giveon, 
E.~Kiritsis, D.~Kutasov, B.~Pioline, E.~Rabinovici, S.~Ribault and V.~Schomerus 
for stimulating discussions.


\end{document}